\newcommand{\be}{\begin{equation}}
\newcommand{\ee}{\end{equation}}
\newcommand{\bea}{\begin{eqnarray}}
\newcommand{\eea}{\end{eqnarray}}
\newcommand{\bean}{\begin{eqnarray*}}

\newcommand{\eean}{\end{eqnarray*}}
\font\upright=cmu10 scaled\magstep1
\font\sans=cmss10
\newcommand{\ssf}{\sans}
\newcommand{\stroke}{\vrule height8pt width0.4pt depth-0.1pt}
\newcommand{\Z}{\hbox{\upright\rlap{\ssf Z}\kern 2.7pt {\ssf Z}}}

\newcommand{\C}{{\rlap{\rlap{C}\kern 3.8pt\stroke}\phantom{C}}}
\newcommand{\R}{\hbox{\upright\rlap{I}\kern 1.7pt R}}
\newcommand{\CP}{\C{\upright\rlap{I}\kern 1.5pt P}}
\newcommand{\PP}{\hbox{\upright\rlap{I}\kern 1.5pt P}}

\newcommand{\identity}{{\upright\rlap{1}\kern 2.0pt 1}}

\newcommand{\pp}{\Delta}
\newcommand{\HH}{\mbox{\hbox{\upright\rlap{I}\kern 1.7pt H}}}

\newcommand{\zb}{{\bar z}}
\newcommand{\fac}{(1+\vert z\vert^2)^2}

\newcommand{\fr}{\frac}
\newcommand{\lm}{\lambda}

\newcommand{\pr}{\partial}
\newcommand{\hs}{\hspace{5mm}}

\newcommand{\dg}{\dagger}

\newcommand{\ve}{\varepsilon}
\newcommand{\acc}{\\[3mm]}

\input{epsf}

\documentstyle[12pt,a4wide]{article}

\newcommand{\news}{\setcounter{equation}{0}}
\begin{document}
\title{\vskip -70pt
\begin{flushright}
{\normalsize UKC/IMS/99/22} \\
\end{flushright}\vskip 50pt
{\bf \large \bf Non-Bogomolny SU(N) BPS Monopoles}\\[30pt]
\author{Theodora Ioannidou and Paul M. Sutcliffe\\[10pt]
\\{\normalsize  {\sl Institute of Mathematics, University of Kent at Canterbury,}}\\
{\normalsize {\sl Canterbury, CT2 7NZ, U.K.}}\\
{\normalsize{\sl Email : T.Ioannidou@ukc.ac.uk}}\\
{\normalsize{\sl Email : P.M.Sutcliffe@ukc.ac.uk}}\\}}
\date{May 1999}
\maketitle

\begin{abstract}
\noindent For $N>2$ we present static monopole solutions of the second order
$SU(N)$ BPS Yang-Mills-Higgs equations which are not solutions of the first order Bogomolny
equations. These spherically symmetric solutions may be
interpreted as monopole anti-monopole configurations and their construction
involves harmonic maps into complex projective spaces.

\end{abstract}

\newpage
\section{Introduction}
\news\ \ \ \ \ \
In this paper we study static monopole solutions of the $SU(N)$
Yang-Mills-Higgs equations in $\R^3,$ in the BPS limit of vanishing
Higgs potential. It is well known that the solutions which correspond to the
global minima of the Yang-Mills-Higgs energy functional are all given by
solutions of the simpler first order Bogomolny equations \cite{Bo}.
However, as proved by Taubes \cite{Ta1} using infinite dimensional Morse theory, there
are more solutions to the Yang-Mills-Higgs equations than just the Bogomolny ones.
These solutions are saddle points of the energy functional and correspond to monopole 
anti-monopole configurations, which have an instability to annihilation.

We construct a class of spherically symmetric $SU(N)$ non-Bogomolny monopoles and calculate some
of their properties, such as magnetic charges and energies. It is interesting that this construction
involves harmonic maps of the plane into \CP$^{N-1}.$  Atiyah \cite{At3} has described
a relationship between instantons in two and four dimensions, and by reduction
a correspondence between $SU(N)$ hyperbolic Bogomolny monopoles and instantons of the
two-dimensional \CP$^{N-1}$ sigma model. 
 Our non-Bogomolny solutions are obtained by using the non-instanton
solutions of the \CP$^{N-1}$ sigma model and so our results appear to suggest that 
Atiyah's connection between Bogomolny monopoles and sigma model instantons may have
 some form of an extension outside the self-dual sector. In any case, it is clear
that we have found some monopole analogues of the sigma model non-instanton solutions.

\section{SU(N) Monopoles}
\news\ \ \ \ \ \
Static monopoles are solutions of the $SU(N)$
Yang-Mills-Higgs equations in $\R^3$ which, in the BPS limit of a massless
Higgs, are derived from the energy functional
\be
E=-\frac{1}{4\pi}\int \mbox{tr}\{(D_i\Phi)^2+\frac{1}{2}F_{ij}^2\}
\ d^3x
\label{energy}
\ee
where $A_i$, for $i=1,2,3$, is the $su(N)$-valued gauge potential, with
field strength $F_{ij}=\partial_iA_j-\partial_jA_i+[A_i,A_j]$, and $\Phi$
is the $su(N)$-valued Higgs field. Variation of the energy (\ref{energy})
gives the second order Yang-Mills-Higgs equations
\be
D_iD_i\Phi=0, \ \ \ D_iF_{ij}=[D_j\Phi,\Phi].
\label{ymh}
\ee
The boundary conditions are that the energy is finite and that, in a chosen
direction (say along the $x_3$-axis), the Higgs field at infinity is a given
constant diagonal matrix, $\Phi(0,0,\infty)=i\Phi_0$ where
\be
\Phi_0=\mbox{diag}(\lambda_1,\lambda_2,...,\lambda_N).
\label{phi0}
\ee
Here we choose the ordering such that
$\lambda_1\ge\lambda_2\ge..\ge\lambda_N$ and because $\Phi\in su(N)$
we have that $\sum_{i=1}^N\lambda_i=0.$ 

At large radius, the magnetic field, $B_k=\frac{1}{2}\ve_{kij}F_{ij},$ 
has the leading order behaviour
\be
B_k
\sim\frac{i\widehat x_k}{2 r^2}G(\widehat x_1,\widehat x_2,\widehat x_3)
\label{defg}
\ee
where the matrix $G$ contains $N-1$ integers, $n_i$, which are the
magnetic charges \cite{GNO} and provide a topological characterization of the
monopole solution. Since we impose the framing condition (\ref{phi0})
along the $x_3$-axis, then along this axis we have that
\be
G_0=G(0,0,1)=\mbox{diag}(n_1,
\,n_2-n_1,\,\dots,\,n_{N-1}-n_{N-2},\,-n_{N-1}).
\label{defg0}
\ee
By completing the square in the energy density (\ref{energy}) the
 Bogomolny bound \cite{Bo} is obtained as
\bea
E&=&-\frac{1}{4\pi}\int \mbox{tr}\{(D_i\Phi\pm B_i)^2\mp 2 B_iD_i\Phi\}\ d^3x
\label{cs}\\
&\ge &
\vert (\lm_1-\lm_2)n_1+(\lm_2-\lm_3)n_2+\dots+(\lm_{N-1}-\lm_N)n_{N-1}\vert.
\label{bb}
\eea
The inequality (\ref{bb}) is obtained by noting that the final term
in (\ref{cs}) can be written as a total derivative and as such can be 
expressed in terms of the magnetic charges and the eigenvalues of the 
Higgs field at infinity.

Clearly, within each magnetic charge sector the minimal energy solutions
are obtained by solving the Bogomolny equations
\be
D_i\Phi=\mp B_i
\label{bog}
\ee
whose solutions saturate the energy bound. The upper sign corresponds
to monopoles, that is, $n_i\ge 0$, whereas choosing the lower sign,
which we shall refer to as the anti-Bogomolny equations, results in 
anti-monopoles with $n_i\le 0$. 

>From the energy bound (\ref{bb}), which
gives the energy for Bogomolny monopoles, it can be seen that 
the difference $m_j=\lm_j-\lm_{j+1}$
determines the mass of the monopole of type $j$, of which there are $n_j$ in
total. 
If the eigenvalues of the Higgs field are not distinct, 
say $\lambda_j=\lambda_{j+1}$, then $m_j=0$, that is,
 the monopole of type $j$ is massless and
 the integer $n_j$ does not appear
in the Bogomolny bound (\ref{bb}). This reflects the fact that $n_j$ is no longer
a topological quantity. In this case the residual symmetry group
is non-abelian, rather than being the maximal torus $U(1)^{N-1}$,
and the integer $n_j$ is not a magnetic charge. To distinguish these
cases such integers are referred to as magnetic weights \cite{GNO}.

Although the global minima of the energy functional (\ref{energy})
are all obtained as Bogomolny monopoles, that is, solutions of
the Bogomolny equations (\ref{bog}), Taubes \cite{Ta1} has proved the existence
of other critical points. In other words, there are solutions of the
full second order Yang-Mills-Higgs equations (\ref{ymh}) which are
not solutions of the first order Bogomolny equations (\ref{bog}).
In the following sections we construct examples of spherically symmetric
 non-Bogomolny monopoles and investigate some of their
properties. In general it is a much more difficult task to 
solve the full Yang-Mills-Higgs equations than the Bogomolny equations,
not just because the equations are second order instead of first order,
but because the Bogomolny equations are integrable and so a variety of
techniques from integrable systems can be applied, whereas this
property is lost for the Yang-Mills-Higgs equations.

\section{The Harmonic Map Ansatz}
\news\ \ \ \ \ \
\label{sec-hma}
The starting point for our investigation is the introduction of the coordinates 
$r,z,\zb$ on $\R^3.$ 
In terms of the usual spherical coordinates
$r,\theta,\varphi$  the Riemann sphere variable is
 $z=e^{i\varphi}\tan(\theta/2).$ Using these coordinates
the Yang-Mills-Higgs equations (\ref{ymh}) take the form
\bea
[D_r\Phi,\Phi]&=&\frac{\fac}{2r^2}(D_zF_{r{\zb}}+D_{\zb} F_{rz})\label{ymh1}\\
D_r(r^2D_r\Phi)&=&-\frac{\fac}{2}(D_zD_{\zb}\Phi+D_{\zb} D_z\Phi)\label{ymh2}\\
\lbrack D_z\Phi,\Phi]+D_r F_{rz}&=&\frac{1}{2r^2}D_z({\fac} F_{z{\zb}})\label{ymh3}
\eea
and the Bogomolny equations (\ref{bog}) become
\be
iD_z\Phi=F_{rz}, \ \ \ iD_r\Phi=\frac{\fac}{2r^2}F_{z\zb}.
\label{bog2}
\ee
>From equation (\ref{defg}) the matrix of magnetic charges, $G$, is given by
\be
G=(1+\vert z\vert^2)^2F_{\zb z}
\label{G}
\ee
where the right hand side of the above is evaluated on the two-sphere at $r=\infty.$

Our ansatz for $SU(N)$ monopoles is to set
\be
\Phi=i\sum_{j=0}^{N-2} h_j(P_j-\frac{1}{N}), \ \ \ 
A_z=\sum_{j=0}^{N-2} g_j[P_j,\partial_z P_j], \ \ \ 
A_r=0
\label{ansatz}
\ee
where $h_j(r),g_j(r)$ are real functions depending only on the radial
coordinate $r$, and $P_j(z,\zb)$ are $N\times N$ hermitian projectors,
that is, $P_j=P_j^\dagger=P_j^2$, which are independent of the radius $r.$
The set of $N-1$ projectors are taken to be orthogonal, so that $P_iP_j=0$
for $i\neq j.$
Note that we are working in a real gauge, so that $A_\zb=-A_z^\dagger$.
In (\ref{ansatz}), and for the remainder of the paper, we drop the
summation convention.

The above ansatz is motivated by our recent study \cite{IS1} of Bogomolny
 monopoles 
and their construction in terms of harmonic maps. In the case of Bogomolny
monopoles it was convenient to a choose a particular complex gauge, but the
existence of this complex gauge choice relies on a solution of the Bogomolny
equations and in general is not valid for the Yang-Mills-Higgs equations. 
However, after converting these Bogomolny solutions to a real gauge they
have the above form, although the ansatz (\ref{ansatz}) is more general.

Substituting the ansatz (\ref{ansatz}) into the Yang-Mills-Higgs equations
one finds that the left hand side of equation (\ref{ymh1}) is identically
zero. This follows from the fact that the projectors are independent of
$r$ and form an orthogonal set. The requirement that the right hand side of
 equation (\ref{ymh1}) is zero gives the following condition
\be
\sum_{j=0}^{N-2} g_j'[P_j,\partial_z\partial_\zb P_j]=0.
\ee
The equation
\be
[P,\partial_z\partial_\zb P]=0
\label{hmap}
\ee
is the harmonic map equation of the two-dimensional \CP$^{N-1}$ sigma model
(see for example ref.\cite{Za}). Thus we take each $P_j$ to be a harmonic
map and then the first Yang-Mills-Higgs equation (\ref{ymh1}) is automatically
satisfied. It is satisfying that the harmonic map equation emerges naturally
from the Yang-Mills-Higgs equations since in the study  of 
Bogomolny monopoles it was found \cite{IS1} to be useful to introduce harmonic maps
but the equations themselves did not appear.

To proceed further we need to briefly recall some results about 
harmonic maps of the two-dimensional \CP$^{N-1}$ sigma model.
See Zakrzewski \cite{Za} for a more detailed account of two-dimensional 
sigma models and their solutions.

If we regard the second order harmonic map equation (\ref{hmap}) as
a lower dimensional analogue of the Yang-Mills-Higgs equations (\ref{ymh}),
then the analogue of the first order Bogomolny equations (\ref{bog}) is
the instanton equation
\be
P\partial_z P=0
\label{inst2}
\ee
whose solutions automatically satisfy the 
 harmonic map equation (\ref{hmap}). The instanton equation (\ref{inst2})
is easy to solve, with the general solution being given by
\be
P(f)=\frac{ff^\dagger}{\vert f\vert^2}
\label{ftop2}
\ee
where $f(z)$ is an $N$-component column vector which is a holomorphic function of $z$ and
 whose degree is equal to the instanton number
 of the sigma model. Another set of solutions are the anti-instantons,
which satisfy the equation $P\partial_\zb P=0$,
and have the same form as the instanton solutions but this time $f$
 is an anti-holomorphic function. For anti-instantons
the sigma model instanton number is minus the degree of $f.$

For $N=2$ these are all the finite action solutions to (\ref{hmap}), but for $N>2$ there are other non-instanton
solutions. These can be obtained from the instanton solutions by a process of
differentiation and Gram-Schmidt orthogonalization. Explicitly, introduce 
 the operator $\pp$ defined  by its action   
on any vector $f\in \C^N$ as
\be
\pp f=\pr_z f- \fr{f \,(f^\dg \,\pr_z f)}{|f|^2}
\ee
and then define further vectors $\pp^k f$ by induction as
$\pp^k f=\pp(\pp^{k-1} f)$.

When calculating with these objects it is useful to be aware of the 
following properties \cite{Za} of
$\pp^k f$ when $f$ is holomorphic:
\begin{eqnarray}
\label{bbb}
&&(\pp^k f)^\dg \,\pp^l f=0, \hs \hs \hs k\neq l\acc
&&\pr_{\bar{z}}\left(\pp^k f\right)=-\pp^{k-1} f \fr{|\pp^k
f|^2}{|\pp^{k-1} f|^2},
\hs \hs
\pr_{z}\left(\fr{\pp^{k-1} f}{|\pp^{k-1} f|^2}\right)=\fr{\pp^k
f}{|\pp^{k-1}f|^2}.
\label{aaa}
\end{eqnarray}
Defining the projectors $P_k$ corresponding to the
family of vectors $\pp^k f$, that is,
\be
P_k=P(\pp^k f), \ \ k=0,..,N-1
\ee
gives our required set of orthogonal harmonic maps.
Since the projectors obtained from this sequence always satisfy
the relation $\sum_{k=0}^{N-1}P_k=1$ and we are going to be taking
arbitrary linear combinations, then we can neglect the final projector
$P_{N-1}.$

Note that applying $\pp$ a total of $N-1$ times to a holomorphic vector gives an anti-holomorphic
vector, so that a further application of $\pp$ gives the zero vector
and hence no corresponding projector. In the \CP$^1$ case the operator $\pp$
converts a holomorphic vector to an anti-holomorphic vector, that is, instantons
to anti-instantons and these are all the solutions in this case.

In order for our ansatz (\ref{ansatz}) to give solutions to the two remaining
Yang-Mills-Higgs equations, (\ref{ymh2}) and (\ref{ymh3}), the harmonic maps
used must have spherical symmetry --- essentially the factors of $\fac$
which appear in the Yang-Mills-Higgs equations must be cancelled.
The required harmonic maps are obtained by applying the above procedure
to the initial holomorphic vector
\be
f=(f_0,...,f_j,...,f_{N-1})^t, \ \ \mbox{where} \ \ f_j=z^j\sqrt{{N-1}\choose j}
\label{smap}
\ee
and ${N-1}\choose j$ denote the binomial coefficients.
For a discussion of the spherical symmetry of these maps see ref.\cite{IS1}.
Here we merely point out that it is at least plausible that the required 
factors do indeed cancel since $\vert f\vert^2=(1+\vert z\vert^2)^{N-1}.$

In the following sections we shall describe the non-Bogomolny monopoles obtained
from our harmonic map ansatz in some
detail for the simplest cases of $SU(3)$ and $SU(4).$ The situation for
general $SU(N)$ will then become clear.

\section{Spherical Monopoles}
\news\ \ \ \ \ \
In dealing with the equations which arise from the harmonic map ansatz (\ref{ansatz})
it is convenient to exchange the profile functions $h_j(r),g_j(r)$ for the
functions $b_j(r),c_j(r)$ which are defined as the following linear combinations
\be
h_j=\sum_{k=j}^{N-2}b_k, \ \ \ c_j=1-g_j-g_{j+1}, \ \ \ \ \mbox{for\ \ } j=0,\ldots,N-2.
\ee
In the above we have defined $g_{N-1}=0.$ Provided the eigenvalues of the Higgs
field at infinity are correctly ordered, as in (\ref{phi0}) (which corresponds to
$b_j(\infty)\ge 0$), then the monopole masses
are simply given by the asymptotic values of the functions $b_j(r)$, that is,
$m_{j}=b_{j-1}(\infty)$ for $j=1,\ldots,N-1.$ Thus if $b_j(\infty)=0$ this signals
a change of symmetry breaking to a non-maximal case. This will be an important point in what
follows.
For the ansatz to be well-defined at the origin the boundary conditions $b_j(0)=0$
and $c_j(0)=1$ for all $j=0,\ldots,N-2$, must be imposed.

\subsection{SU(2)}
\news\ \ \ \ \ \
As we have mentioned in section \ref{sec-hma}, 
there are no non-instanton solutions of the \CP$^1$ sigma model and hence we
cannot employ our ansatz to obtain non-Bogomolny monopoles for gauge group $SU(2).$
There are only two profile functions $b_0,c_0$, and the only solution is the standard
spherically symmetric Bogomolny 1-monopole.
The non-Bogomolny $SU(2)$ monopole of Taubes \cite{Ta1} is not spherically symmetric,
it has only an axial symmetry, and thus does not fall into the class of solutions which
we obtain here.

\subsection{SU(3)}
\news\ \ \ \ \ \
For $N=3$ there are four profile functions, $b_0,b_1,c_0,c_1$, and our ansatz (\ref{ansatz})
reduces the Yang-Mills-Higgs equations to the following set of second order nonlinear ordinary
differential equations
\bea
(b_j'r^2)'&=&2(2b_jc_j^2-b_kc_k^2)\cr
r^2c_j''&=&c_j(2c_j^2-c_k^2-1+b_j^2r^2).
\label{su3}
\eea
Here the indices are chosen from the set $\{0,1\}$ and $k\neq j.$ Recall that the summation
convention is no longer used in this paper. It is immediately clear that there is a symmetry
under the interchange of indices, $0\leftrightarrow 1$, when applied simultaneously to both
the $b_j$ and $c_j$ functions;  we shall make use of this symmetry later.

The corresponding energy, (\ref{energy}), is given by
\be
E=2\int_0^\infty \frac{r^2}{3}(b_0'^2+b_1'^2+b_0'b_1')
+c_0'^2+c_1'^2+c_0^2b_0^2+c_1^2b_1^2+\frac{1}{2r^2}
((1-c_0^2)^2+(1-c_1^2)^2+(c_0^2-c_1^2)^2) \ dr.
\label{su3energy}
\ee
>From this expression it can be seen that the energy is finite providing the functions
approach their asymptotic values at least as fast as $1/r$, and if in addition
the constraints that $c_j(\infty)b_j(\infty)=0$ are imposed for $j=0,1.$

Before studying the second order equations (\ref{su3}) it is first useful to examine the 
first order Bogomolny equations, which in this formalism become
\bea
r^2b_j'&=&-(2c_j^2-c_k^2-1)\cr
c_j'&=&-c_jb_j
\label{su3bog}
\eea
where the notation is as above. Integrating the last equation gives the asymptotic
behaviour for $c_j$ as
\be
c_j\sim \exp(-rb_j(\infty)+O(1)).
\ee
Now since $c_j$ must be finite as $r\rightarrow\infty$
this gives that $b_j(\infty)\ge 0.$ Thus we can characterize Bogomolny monopoles
by the fact that the asymptotic values of the $b_j$'s are all non-negative. 
If we consider the anti-Bogomolny equations then they are given by equations
(\ref{su3bog}) but in which the minus signs are removed from the right hand side of the
equations. In this case the requirement that $c_j(\infty)$ is finite implies that
$b_j(\infty)\le 0.$ Thus we see that Bogomolny monopoles have the
property that all the asymptotic values of the $b_j$'s have the same sign, positive
for monopoles and negative for anti-monopoles. It is then natural to look for non-Bogomolny
solutions of the second order equations in which the asymptotic values $b_0(\infty)$
and $b_1(\infty)$ have opposite sign, and to interpret these as monopole anti-monopole solutions.

In order to read off the properties of a given solution we need to compute the 
Higgs field and magnetic charge matrix at ${\bf x}=(0,0,\infty)$  (which corresponds to the
direction $z=0$). Explicitly, these are given by
\bea
\Phi_0&=&\frac{1}{3}\mbox{diag}(2b_0+b_1,-b_0+b_1,-b_0-2b_1)\label{q1}\\
G_0&=&\mbox{diag}(2(1-c_0^2),2(c_0^2-c_1^2),2(c_1^2-1)).\label{q2}
\eea
As an example, consider the Bogomolny monopole with maximal symmetry breaking
and equal monopole masses given by $b_0(\infty)=b_1(\infty)=2.$ Then
$\Phi_0=\mbox{diag}(2,0,-2)$ and the boundary conditions force that $c_0(\infty)=c_1(\infty)=0$
so that $G_0=\mbox{diag}(2,0,-2).$ Comparing with equation (\ref{defg0}) we see that
the magnetic charges are $(n_1,n_2)=(2,2).$ For Bogomolny monopoles the solutions can be
obtained explicitly and the monopole charges are understood in terms of the degrees of the
harmonic map projectors from which they are constructed \cite{IS1}.
At first site it might appear from (\ref{q2}) that the magnetic charges are determined
 only by the boundary values $c_j(\infty)$ and are independent of the values of $b_j(\infty).$
However, this naive view is incorrect as is easily seen by considering
the simple case of the anti-Bogomolny solution with $b_0(\infty)=b_1(\infty)=-2.$
Again the boundary conditions imply that $c_0(\infty)=c_1(\infty)=0$
and hence we obtain the same matrix $G_0=\mbox{diag}(2,0,-2).$ 
But now we must be aware
that in this case we have $\Phi_0=\mbox{diag}(-2,0,2)$ so that the entries
are not correctly ordered from the largest to the smallest. A constant gauge
transformation permutes the entries to obtain $\Phi_0=\mbox{diag}(2,0,-2)$
but this acts in the same way on the magnetic charge matrix so that after this
gauge transformation we are left with the charge matrix $G_0=\mbox{diag}(-2,0,2).$
Now that $\Phi_0$ has the correct order we can compare this charge matrix with
equation (\ref{defg0}) and conclude that  $(n_1,n_2)=(-2,-2).$ Although this example
of computing the magnetic charges is trivial it illustrates the important point
that the asymptotic values $b_j(\infty)$ are required in order to determine the
magnetic charges. We shall see more interesting consequences of this fact in what follows.

For the moment we shall consider the case 
for which $b_j(\infty)\neq 0$, 
so that the boundary conditions are $c_j(\infty)=0$ for $j=0,1.$
The Bogomolny equations (\ref{su3bog}) are integrable and allow explicit solutions
to be found for any choice of the positive parameters $b_0(\infty),b_1(\infty)$,
which give the monopole masses $m_1,m_2.$ However, it seems unlikely that explicit 
non-Bogomolny solutions to the second order equations (\ref{su3}) can be found in closed
form. Therefore we resort to a numerical solution of these equations. We apply a gradient
flow algorithm with a finite difference scheme to compute the solution with a given set 
of boundary values $b_0(\infty),b_1(\infty).$ For all choices of these parameters we were
able to find a numerical solution. As a test on the accuracy of the code we computed
the charge $(2,2)$ Bogomolny solution with $b_0(\infty)=b_1(\infty)=2.$ In this case, 
since the monopole masses are equal, $m_1=m_2=2$, and the total number of monopoles is
four then the energy is $E=8.$ This value of the energy was obtained from our numerical
code to within an accuracy of three decimal places. 

In order to consider the non-Bogomolny analogue of this solution we want to
fix the monopole masses in the same way as $m_1=m_2=2.$ Thus the eigenvalues of
 $\Phi_0$ must again be $0$ and $\pm 2$, though this time their order will not be correct.
For example, consider the choice of ordering $\Phi_0=\mbox{diag}(0,2,-2),$  which by 
(\ref{q1}) corresponds to the boundary values $b_0(\infty)=-2$ and $b_1(\infty)=4.$
Thus, since $b_0(\infty)$ and $b_1(\infty)$ have opposite sign, this gives a non-Bogomolny
solution. In Figure 1.1 we plot the functions $b_0,b_1,c_0,c_1$ obtained from the
numerical solution in this case. In Figure 1.2 we plot the energy density of this solution
(solid line) and the energy density of the corresponding Bogomolny solution (dashed line).
Note the dip in the energy density at the origin for the non-Bogomolny solution, so that
some energy density isosurfaces will be shell-like. We compute the energy of this
solution to be $E=9.0,$ so that it is larger than the Bogomolny solution. As discussed
above, the entries of the magnetic charge matrix must also be permuted (in accordance
with the permutation of the entries in the Higgs field to obtain the correct ordering)
and this results in $G_0=\mbox{diag}(0,2,-2).$ Comparison with (\ref{defg0}) then
gives the charges as $(n_1,n_2)=(0,2).$ Clearly the energy of this solution has little
to do with the Bogomolny bound (\ref{bb}), and it would be nice to understand its value.
With this aim in mind we now attempt some phenomenology to interpret the charge and
energy of this monopole. 

As mentioned in section \ref{sec-hma} the projectors used in the harmonic map ansatz
have a sigma model interpretation in terms of instanton anti-instanton configurations.
However, the Bogomolny monopole solutions clearly have no anti-monopoles, so it
appears that in this case the profile functions are such that the monopole does not
see any anti-soliton content. Nevertheless, when the profile functions are modified
to a non-Bogomolny solution some of the anti-soliton content becomes visible --- we have
already seen a signature for this in terms of the signs of the asymptotic values of
the profile functions. This suggests that we should think of the charge $(0,2)$ solution
as the composite $(0,2)=(+2-2,+2-0)$, where the plus signs denote monopoles and the minus
signs anti-monopoles (the positive monopole content is taken from the Bogomolny
solution). With this interpretation the $(0,2)$ solution contains two monopoles and
two monopole anti-monopole pairs. Since the energy of the solution is $E=9.0$ and the
monopole mass is 2, this phenomenology gives an approximate value for the energy
of a monopole anti-monopole pair as $E_{m\bar m}=2.5.$ This value is at least reasonable,
since (having normalized the monopole mass to 2) the energy of a monopole anti-monopole
pair should be something less than 4; the precise value depends on the details of the
monopole anti-monopole interaction. Some non-trivial tests of the above interpretation
will arise later when we consider $SU(4)$ monopoles. Note that although we know that these
solutions are not global minima and we expect them to be unstable (to annihilation of
the monopole anti-monopole pairs) we have not proved that they are unstable. 
A stability analysis would need to be undertaken to prove that the solutions are
not local minima, though it would be extremely surprising if they were stable.

In $SU(3)$ there are six possible ways to order the eigenvalues of the Higgs field and the
corresponding monopole charges are $\pm(2,2),(0,\pm2),(\pm2,0)$, thus all the other
monopole solutions are trivially related to the two examples we have discussed.

Let us now turn to minimal symmetry breaking, which we may take to be given by
$\Phi_0=\mbox{diag}(1,-\frac{1}{2},-\frac{1}{2})$. The Bogomolny monopole has boundary
values given by $b_0(\infty)=\frac{3}{2}$ and $b_1(\infty)=0.$ Since the second 
profile function vanishes at infinity then the boundary conditions now allow that
$c_1(\infty)\neq 0.$ As this is a Bogomolny solution then it can be found explicitly
and the appropriate boundary condition turns out to be \cite{IS1}
$c_1(\infty)=\frac{1}{\sqrt{2}}$, which gives the charge $(2,[1]).$ Here the notation
is that magnetic weights are denoted by square brackets. The monopole masses are
$m_1=\frac{3}{2},m_2=0$, so the energy is $E=3.$

Recall that the Bogomolny solutions have the property that all the $b_j$'s at
infinity have the same sign, whereas for non-Bogomolny solutions this is not the case.
Thus one may wonder whether these two types of solution are smoothly connected
as the values of (one or more) $b_j(\infty)$ are varied to change sign. The answer
is that they are not, since this smooth variation must pass through the point
$b_j(\infty)=0$ where (as we see above) the symmetry breaking pattern changes
and the boundary condition on  $c_j(\infty)$ suffers a discontinuous jump.

For Bogomolny monopoles a non-maximal symmetry breaking pattern
requires the vanishing of at least one $b_j(\infty)$, but for non-Bogomolny
monopoles this is not the case. For example, permuting the eigenvalues
into the order $\Phi_0=\mbox{diag}(-\frac{1}{2},1,-\frac{1}{2})$ corresponds to the
choice $b_0(\infty)=-b_1(\infty)=-\frac{3}{2}$ and hence the charge $(0,[2]).$
The symmetry of equations (\ref{su3}) together with the symmetry of the boundary
conditions in this case force the reduction $c_0=c_1$ and $b_0=-b_1.$
The minimally broken $SU(3)$ non-Bogomolny monopole that arises from this specific
reduction has been obtained previously by Burzlaff \cite{Bu}, using a
hedgehog-like ansatz and a  group theoretic approach in which the gauge potential
 involves the principle $SU(2)$
triplet in $SU(3)$ but the Higgs field involves the associated 5-plet. 
Burzlaff \cite{Bu} proved the existence of a solution to the equations which result
in this case.  An important feature of the proof is the fact that the equations arise
as the variation of an energy functional and therefore it only remains to show that
a minimizer exists, for which standard methods can be employed. In our general case
we have the variational formulation which comes from the Yang-Mills-Higgs energy
and so it should be possible to use similar techniques to prove the existence of
solutions. In this paper we are content with numerical solutions of the profile function
equations and these are shown in Figure 1.3 for this case of charge $(0,[2]).$
In Figure 1.4 we plot the energy density for this solution (solid line) and the
energy density of the corresponding Bogomolny charge $(2,[1])$ solution (dashed line). 
Again note
the dip in energy density near the origin of the non-Bogomolny solution. The energy
of the non-Bogomolny solution is found to be $E=4.3$, which should be compared with
the energy $E=3$ of the Bogomolny solution.

\subsection{SU(4)}
\news\ \ \ \ \ \
For $N=4$ there are six profile functions, $b_j,c_j$, $j=0,1,2$, and 
the Yang-Mills-Higgs equations reduce to
\bea
(r^2b_0')'&=&6c_0^2b_0-4c_1^2b_1\cr
(r^2b_1')'&=&8c_1^2b_1-3c_0^2b_0-3c_2^2b_2\cr
(r^2b_2')'&=&6c_2^2b_2-4c_1^2b_1\cr
r^2c_0''&=&c_0(3c_0^2-2c_1^2-1+b_0^2r^2)\cr
r^2c_1''&=&c_1(4c_1^2-\frac{3}{2}c_0^2-\frac{3}{2}c_1^2-1+b_1^2r^2)\cr
r^2c_2''&=&c_2(3c_2^2-2c_1^2-1+b_2^2r^2).
\eea
If we regard the indices on the profile functions as labelling sites on a linear lattice
then we see that the equations involve only a nearest neighbour coupling. Also, there
is again a symmetry of reflecting the lattice about its midpoint, which in this case
is the interchange $0\leftrightarrow 2.$ 

A consideration of the associated energy shows that the boundary conditions again
require that $c_j(\infty)b_j(\infty)=0$, for all $j.$ In this section we shall mainly
be concerned with maximal symmetry breaking, in which case $c_j(\infty)=0.$

The eigenvalues and charges are given by
\bea
\Phi_0=\frac{1}{4}\mbox{diag}(3b_0+2b_1+b_2,2b_1+b_2-b_0,b_2-b_0-2b_1,-b_0-2b_1-3b_2)
 \label{qq1}\\
G_0=\mbox{diag}(3(1-c_0^2),(1+3c_0^2-4c_1^2),
-(1+3c_2^2-4c_1^2),-3(1-c_2^2)). \label{qq2}
\eea
Normalizing the monopole mass to two, the Bogomolny solution with
 $b_0(\infty)=b_1(\infty)=b_2(\infty)=2$ has $\Phi_0=\mbox{diag}(3,1,-1,-3).$
>From (\ref{qq2}) this solution has $G_0=\mbox{diag}(3,1,-1,-3)$ and hence by comparison with
(\ref{defg0}) the charge is $(3,4,3),$ with corresponding energy $E=20.$

There are 24 different orderings of the eigenvalues $\pm3,\pm1$ and of these there
are 8 which give fundamentally different monopoles.
 In Table 1 we list the values of $b_j(\infty)$
and the corresponding magnetic charges.
 As in the $SU(3)$ case, the magnetic charges are computed by first finding the
permutation required so that $\Phi_0$ has the correct ordering, that is,
 $\Phi_0=\mbox{diag}(3,1,-1,-3)$, then applying this permutation
to the elements of $G_0=\mbox{diag}(3,1,-1,-3)$ which can then be compared with
the definition (\ref{defg0}).\\

\begin{center}
\begin{tabular}{|c|c|c|c|c|c|}
\hline
$b_0(\infty)$&$b_1(\infty)$&$b_2(\infty)$ & $(n_1,n_2,n_3)$ & $E$ & $\widetilde E$\\
\hline
2&2&2&(3,4,3)&20.0&20.0\\
2&4&-2&(3,4,1)&21.6&21.0\\
4&-2&4&(3,2,3)&22.0&21.0\\
-2&6&-2&(1,4,1)&22.9&22.0\\
6&-2&-2&(3,0,-1)&22.9&25.5\\
4&2&-4&(3,0,1)&23.3&23.0\\
6&-4&2&(3,2,-1)&24.1&24.5\\
4&-6&4&(-1,2,-1)&27.0&28.0\\
\hline
\end{tabular}

{\bf Table 1} : Boundary conditions, charges and energies for $SU(4)$ monopoles.
\ \\ 
\end{center}

In Table 1 we also list the computed energies, $E$, of the solutions and the approximate
values, $\widetilde E$, calculated using the monopole anti-monopole interpretation
discussed in the previous section. Thus for example, we write $(3,0,1)=(3-0,4-4,3-2)$,
where the positive monopole content is taken from the Bogomolny solution. The interpretation
of this solution is therefore that it contains six monopole anti-monopole pairs and
four monopoles. The approximate energy is then
 $\widetilde E=4\times 2+6E_{m\bar m}=23,$ where we have used the result of the previous section
that $E_{m\bar m}=2.5.$ The true energy of this solution is $E=23.3$ which is in good agreement
with the approximate value. A glance at Table 1 reveals that the approximate energies are in
reasonable agreement with the calculated values, which adds support to our interpretation of
the solutions in terms of monopole anti-monopole configurations. Only one case, the charge
$(3,0,-1)$ solution, shows a significant discrepancy and it is easy to imagine that the
 finer details of the arrangement of the monopoles/anti-monopoles needs to be taken into account
to get a better agreement. Indeed the surprising fact is that our naive counting 
produces results which are remarkably accurate.

We have only discussed the case of maximal symmetry breaking in this section but
solutions for all symmetry breaking patterns can also be found. As an example,
the case of minimal symmetry breaking can be obtained by exploiting the symmetry
of the equations to set $b_0=-b_2$, $b_1=0$, $c_0=c_2$, $c_1=0.$ This gives the
$SU(4)$ analogue of Burzlaff's $SU(3)$ solution, and indeed the equations
in the two cases have the same structure.

\section{Conclusion}
\news\ \ \ \ \ \
In this paper we have presented some static spherically symmetric monopole solutions of the
$SU(N)$ Yang-Mills-Higgs equations which are not solutions of the Bogomolny equation.
Some of the properties of these solutions have been calculated and their interpretation
in terms of monopole anti-monopole configurations discussed. A crucial tool in the
investigation was the connection with the two-dimensional \CP$^{N-1}$ sigma model, and
indeed our solutions may be considered to be the monopole analogues of the
sigma model non-instanton solutions. Finally, it is perhaps interesting to note
that in the context of string theory Sen \cite{Se1} has revealed new perspectives on D-branes
and other aspects of string theory by consideration of the dynamics of 
 unstable D-brane anti-D-brane configurations. In particular D-branes themselves
appear as topological defects in the worldvolume of higher dimensional unstable brane 
configurations. It remains to be seen whether
monopole anti-monopole solutions have any important role to play in this light, 
but if so then our solutions
are perhaps the simplest examples upon which any further studies might be based.\\

\section*{Acknowledgements}
\news\ \ \ \ \ \
Many thanks to Conor Houghton, Niall Mackay, Nick Manton and Wojtek Zakrzewski
for useful discussions. PMS acknowledges the EPSRC for an Advanced Fellowship
and the grant GR/L88320.\\

\ \\

\section*{Figure Captions}
\news\ \ \ \ \ \
Fig 1.1. The profile functions for the $SU(3)$ non-Bogomolny monopole
with maximal symmetry breaking.

Fig 1.2. Energy density for $SU(3)$ monopoles with maximal symmetry breaking;
non-Bogomolny solution (solid line); Bogomolny solution (dashed line).

Fig 1.3. The profile functions for the $SU(3)$ non-Bogomolny monopole
with minimal symmetry breaking.

Fig 1.4. Energy density for $SU(3)$ monopoles with minimal
 symmetry breaking;
non-Bogomolny solution (solid line); Bogomolny solution (dashed line).

\newpage

\begin{figure}[ht]
\begin{center}
\leavevmode
\epsfxsize=15cm\epsffile{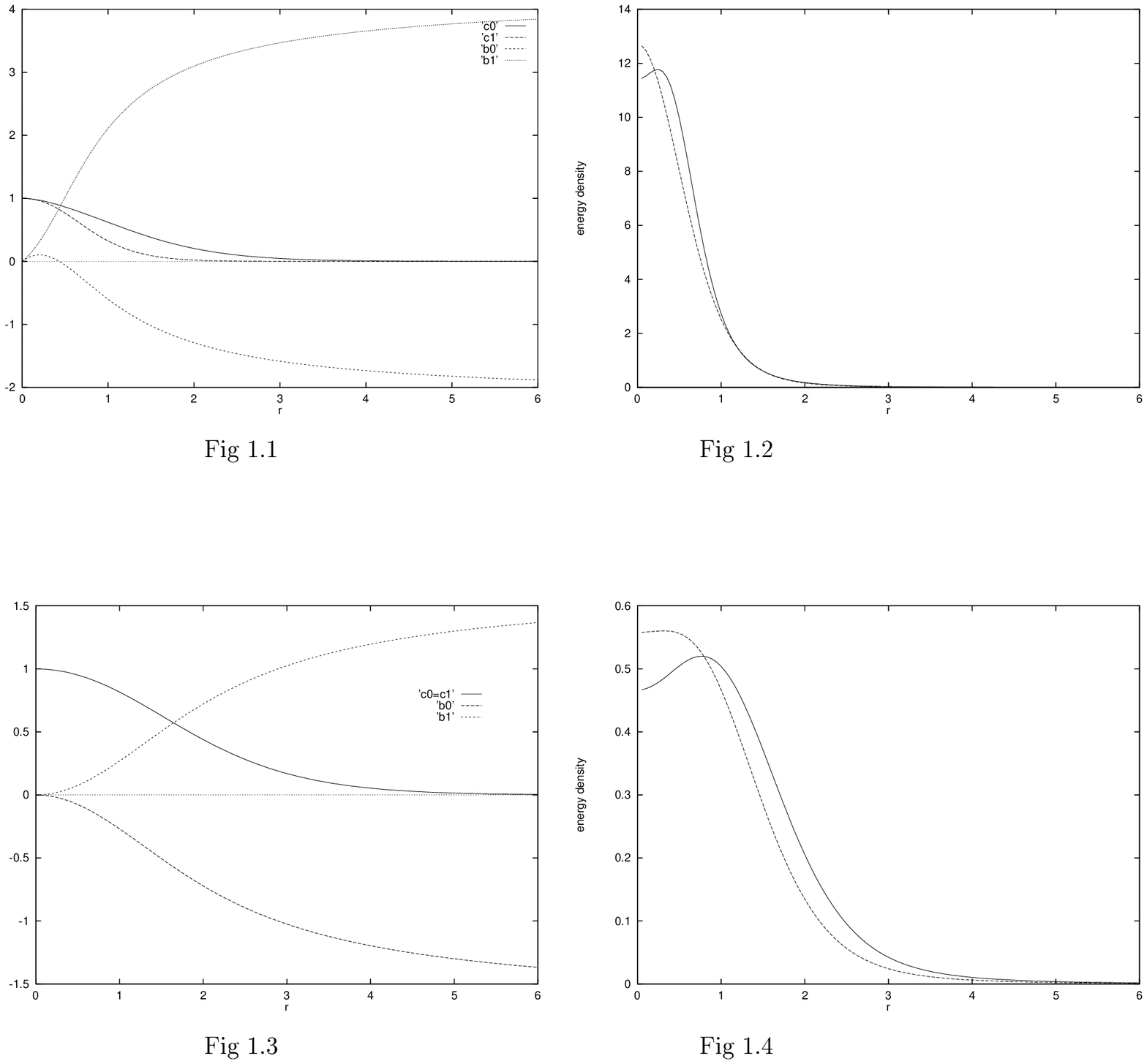}
\caption{ }
\end{center}
\end{figure}


\newpage

\begin{thebibliography}{99}

\bibitem{At3} M.F. Atiyah, Commun. Math. Phys. 93, 437 (1984).

\bibitem{Bo} E.B. Bogomolny, Sov. J. Nucl. Phys. 24, 449 (1976).

\bibitem{Bu} J. Burzlaff, Phys. Rev. D 23, 1329 (1981); Acta Phys. Austriaca Suppl. 1
(1982); Czech. J. Phys. B 32, 624 (1982).

\bibitem{GNO} P. Goddard, J. Nuyts and D. Olive, Nucl. Phys. B 125, 1, (1977).

\bibitem{IS1} T. Ioannidou and P.M. Sutcliffe, \lq{\sl Monopoles and Harmonic Maps}\rq, 
hep-th/9903183.

\bibitem{Se1} A. Sen, hep-th/9803194/9805019/9805170/9808141.

\bibitem{Ta1} C.H. Taubes, Commun. Math. Phys. 86, 257 \& 299 (1982).

\bibitem{Za}
W. J. Zakrzewski, {\it Low dimensional sigma models} (IOP, 1989).

\end{thebibliography}
\end{document}